\documentclass[aps,prd,superscriptaddress,nofootinbib]{revtex4}
\usepackage{amsmath, amsthm, amsfonts}
\usepackage{graphicx}
\usepackage{natbib}

\begin{document}
\title{Calculation of the critical overdensity in the
 spherical-collapse approximation}
\author{D. Herrera, I. Waga and S.E. Jor\'as}
\affiliation{Instituto de F\'\i sica, Universidade
Federal do Rio de Janeiro\\{C. P. 68528, CEP 21941-972, Rio de Janeiro, RJ,
Brazil}}

\begin{abstract}
Critical overdensity $\delta_c$ is a key concept in estimating the number count of halos for different redshift and halo-mass bins, and therefore, it is a powerful tool to compare cosmological models to observations. 
There are currently two different prescriptions in the literature for its calculation, namely, the differential-radius and the constant-infinity methods. In this work we show that the latter yields precise results {\it only} if we are careful in the definition of the so-called numerical infinities.
Although the subtleties we point out are crucial ingredients for an accurate determination of $\delta_c$ both in general relativity and in any other gravity theory, we focus on $f(R)$ modified-gravity models in the metric approach; in particular, we use the so-called large ($F=1/3$) and small-field ($F=0$) limits.
For both of them, we calculate the relative errors (between our method and the others) in the critical density $\delta_c$, in the comoving number density of halos per logarithmic mass interval $n_{\ln M}$ and in the number of clusters at a given redshift in a given mass bin $N_{\rm bin}$, as functions of the redshift. We have also derived an analytical expression for the density contrast in the linear regime as a function of the collapse redshift $z_c$ and $\Omega_{m0}$ for any $F$.

%Although the former two show discrepancies among the different methods used in their calculations, the latter does not.
\end{abstract}

\maketitle

%%%%%%%%%%%%%%%%%%%%%%%%%%%%%%%%%%%%%%%%%%%%%%%%%
\section{Introduction}
 
Since the discovery that the expansion of the Universe is speeding up in $1998$ \cite{ref4,ref5}, several attempts have been made to understand the physical mechanism responsible for this cosmic acceleration. One possibility is that an exotic new component with negative pressure (dubbed dark energy) would be responsible for it. The simplest dark energy candidate is the cosmological constant ($\Lambda$). As an alternative to dark energy, one considers modification of general relativity (GR). The simplest modified-gravity candidate is the so-called $f(R)$  gravity in which the Lagrangian density  ${\cal L} = R + f(R)$ is a nonlinear function of the Ricci scalar $R$. Here, we focus on the metric approach.

There is a general belief (see, for instance, Ref. \cite{Hurtado}) that cosmological kinematical tests (those that depend only on the background evolution) alone cannot distinguish between the standard cosmological constant model ($\Lambda$CDM) and any viable $f(R)$ gravity theory. Indeed, one of the strongest constraints  comes from the growth of perturbations in the nonlinear regime \cite{Schmidt, Santos}. In the linear regime, on the other hand, only weak constraints are obtained \cite{ODweyer}.

In GR, Birkhoff's theorem holds, and in the spherical collapse (SC) approximation, an initial top-hat density profile keeps being a top hat. That straightforwardness enables analytical results in the Einstein-de Sitter (EdS) background ($\Omega_m=1$) ---and it is therefore the standard benchmark for all more realistic initial conditions. In modified theories, however, the fifth force mediated by the new scalar degree of freedom---the scalaron \cite{Starobinsky}---and the so-called chameleon mechanism \cite{camaleon} play a crucial role. 
%%%V2
Indeed, the chameleon mechanism is a key ingredient to hide the fifth-force effects in high-density environments such as the Solar System and at Galactic scales. 
Consequently, 
%%%Indeed, 
the validity of the SC approximation itself has been the subject of a large dispute in the current literature. Borisov {\it et al.} \cite{f1} numerically solved the full modified gravity equations for the model proposed by Hu and Sawicki \cite{Hu} and found that an initial top-hat profile develops shell crossing during its evolution, and therefore, its shape changes. An improvement of the SC numerical calculation for again the same $f(R)$ model is found in Ref. \cite{koop}, using as an initial condition the average density profile around a density peak. Using the results for the SC found in Ref. \cite{camaleon}, Lombriser {\it et al.} \cite{Lombriser} and subsequently Cataneo {\it et al.} \cite{cataneo} have taken into account the chameleon suppression of modifications in high-density regions. 

Precisely in order to circumvent such problems and to gain some insight on the role played by the chameleon mechanism, we work in the so-called large- and small-field limits (see Sec \ref{review}).  Indeed, such an approach has been proven effective before: In Refs.~\cite{ref36,ferrano}, the density profiles and the linear bias of the cluster halos were determined in such limits, showing good agreement with N-body simulations. 

The key quantity in the SC is the critical overdensity $\delta_c(z_c)$ at a given collapse redshift $z_c$. It is defined as the final value (i.e, at redshift $z_c$) of the  linear evolution of a given spherical top-hat initial perturbation that actually collapses at $z_c$ according to the full nonlinear equations. Theoretically, the latter value should be infinite, but in practice one has to deal with numerical infinities, and here, is where lies a potential problem. As we see later on, different numerical infinities give rise to different results for $\delta_c(z)$. The main goal of this work is to discuss and carefully quantify these differences.

This paper is structured as follows. Section \ref{review} reviews the basic equations that describe the SC model in $f(R)$ theories. In  Sec. \ref{methods} we review the current techniques to calculate the critical density contrast and point out the delicate step in the constant-infinity method. In Sec. \ref{conclusion}  we compare and discuss the results. In particular, we show the relative errors between our method and the others, in the case of  $f(R)$ theories, in both the so-called small-field limit ($F=0$) and the large-field limit ($F=1/3$), when calculating the critical density $\delta_c$, the comoving number density of halos per logarithmic mass interval $n_{\ln M}$, and the number of clusters at a given redshift in different mass bins $N_{\rm bin}$. Throughout this paper, we assume the background is given by the standard $\Lambda$CDM model, since we assume there is no observable difference between that and the actual viable $f(R)$ evolution. In the Appendix we present the evolution of the density contrast in the linear regime for arbitrary $\Omega_{m0}$.

%%%%%%%%%%%%%%%%%%%%%%%%%%%%%%%%%%%%%%%%%%%%%%%%%%%% 
\section{Spherical collapse in $f(R)$ theories}
\label{review}

In $f(R)$ theories the Einstein-Hilbert action is modified to
\begin{equation}\label{1}
S=  \int  d^{4}x \sqrt{-g}\left[\frac{R+ f(R)}{2\kappa} + {\cal L}_{m}\right],
\end{equation}
where ${\cal L}_{m}$ is the Lagrangian of the ordinary matter, $\kappa \equiv 4\pi G$, and throughout this paper, we use $c=\hbar=1$. Variation of Eq. (\ref{1}) with respect to the metric $g^{\mu\nu}$ yields the modified Einstein equations:
\begin{equation}\label{2}
G_{\mu\nu}+f_{R}\,R_{\mu\nu} - \frac{1}{2}f\,g_{\mu\nu}- [\nabla_{\mu}\nabla_{\nu}-g_{\mu\nu}\square]f_{R}=\kappa T_{\mu\nu},
\end{equation}
where $G_{\mu\nu}$ is the Einstein tensor and $f_{R}\equiv \frac{df}{dR}$. Taking the trace of Eq. (\ref{2}), one gets
\begin{equation}\label{3}
f_{R}\,R-2f+3\,\square f_{R}-R=\kappa T,
\end{equation}
where $T$ is the trace of the energy-momentum tensor $T_{\mu\nu}$. From the latter, equation one can see that $f_R$ represents an extra degree of freedom. Indeed, one can show \cite{m18} that upon a change to Einstein's frame, one recovers standard GR plus an extra scalar field.

The SC model considers a homogeneous and isotropic region (a top-hat profile) with density $\rho(t)=\bar{\rho}(t)+\delta \rho(t)$, where $\bar{\rho}$  is the background fluid density. We suppose that this region contains only nonrelativistic matter (both the pressure $p_m$ and the effective sound speed $c_{\rm eff}^2$ are negligible). 
Such a region can be described as a perturbation in an otherwise homogeneous universe with density $\bar\rho (t)$, scale factor $a(t)$, and Hubble parameter $H\equiv \dot a /a$, whose metric is given by 
\begin{equation}\label{7}
ds^{2}=-(1+2\phi)dt^{2} + a^{2}(t)(1+2\psi)\delta_{ij}dx^{i}dx^{j}.
\end{equation}
In GR, in the considered case (since the anisotropic stress vanishes), the gravitational potential $\phi$ would be equal to the negative of the second potential ($\phi +\psi=0$). In modified theories, however, the extra scalar field acts as a source of the deviation between them. 

The nonlinear continuity and Euler equations \cite{Peebles} for the density contrast $\delta(t)\equiv\frac{\delta \rho}{\bar{\rho}}$  and velocity-field perturbation $\vec v$ are, respectively,
\begin{equation}\label{4}
\dot{\delta}+\frac{1}{a}(1+\delta)\vec{\nabla}\cdot \vec{v}=0 \quad {\rm and}
\end{equation}
\begin{equation}\label{5}
\dot{\vec{v}}+\frac{1}{a}(\vec{v}\cdot \vec{\nabla})\vec{v}+H\vec{v}= -\frac{1}{a}\vec{\nabla}\phi,
\end{equation}
in comoving spatial coordinates. Combining Eqs.~(\ref{4}) and (\ref{5}), one obtains a second-order differential equation for $\delta$:
\begin{equation}\label{6}
\ddot{\delta}+2H\dot{\delta}-\frac{4}{3}\frac{\dot{\delta}^{2}}{(1+\delta)}= \frac{1}{a^{2}}\nabla^{2}\phi \, (1+\delta).
\end{equation}

When gravity is modeled by $f(R)$ theories, the potential $\phi$ is modified accordingly, as follows. A perturbation in the matter density produces a perturbation in the metric $g_{\mu\nu}$, which can be translated into a perturbation in the Ricci scalar $R$. 
Using  Eqs.~(\ref{2}) and (\ref{7}), one gets the equation for the modified potential $\phi$:
\begin{equation}\label{8}
\nabla^{2}\phi=\frac{16 \pi G}{3}a^{2}\delta \rho_{m}-\frac{a^{2}}{6}\delta R(f_{R}),
\end{equation}
where $\delta R\equiv R-\bar{R}$ and $\bar R$ is the background Ricci scalar. Accordingly, the function  $f_{R}(\bar{R})\equiv \frac{df}{dR}(\bar{R})$ is perturbed by $\delta f_{R}\equiv f_{R}(R)-f_{R}(\bar{R})$. Using (\ref{3}), we get
\begin{equation}\label{9}
\nabla^{2}\delta f_{R}= \frac{a^{2}}{3}[\delta R (f_{R})-8\pi G \delta \rho_{m}].
\end{equation}

To obtain Eqs.~(\ref{8}) and (\ref{9}), we have considered  $| f_{R}(\bar{R})| \ll 1$ and the quasi-static approximation $\dot{f}_{R}(\bar{R}) \ll |\vec\nabla f_{R}(\bar{R})|$. The former condition indicates that the background is similar to $\Lambda$CDM, while the latter assumes that the time scale of the collapse is much smaller than the time scale of the expansion of the Universe. Therefore, any time variation of the (background) scalar field  $f_{R}$ is negligible in a typical time scale of collapse.  
%%%V2
Such an approximation is equivalent to focus on the evolution of the perturbations inside the Hubble radius when the background evolution is close to $\Lambda$CDM ($f_{R}\ll 1$), as has been shown in 
Ref.~\cite{ref35}. However, in Ref.~\cite{PhysRevD.89.084023}, it was shown that the deviation in the global matter power spectrum between static and nonstatic simulations is only $~0.2\%$. Therefore, although in principle the static approximation is not supposed to be accurate, the corrections are actually small.
%%% 

Two opposite regimes, inherent to any $f(R)$, can be represented by a single factor $F$ \cite{ref36}. When the curvature in the spherical region is large, the fluctuations of the field $\delta f_{R}$ are very small, so that the Laplacian in Eq. (\ref{9}) can be neglected, yielding  $\delta R\approx 8\pi G \delta \rho_{m}$ and the usual Poisson equation is recovered; this defines the so-called small-field limit. 
On the other hand, if the curvature $R$ in the spherical region is similar to the background curvature $\bar {R}$, its fluctuation $\delta R$ is small. Thus, it can be neglected in Eq. (\ref{8}), which increases the gravitational potential by a global $4/3$ factor on the right-hand side of that equation. This regime is the so-called large-field limit, i.e., $| \delta f_{R}|\sim | \phi |$.
In the former case, gravity is not modified due to the chameleon effect; the opposite situation occurs in the large-field limit where gravity is strengthened, becoming more attractive.

For both the limits above, Eq. (\ref {6}) can be cast as
\begin{equation}\label{10}
\ddot{\delta}+2H\dot{\delta}-\frac{4}{3}\frac{\dot{\delta}^{2}}{(1+\delta)}=\frac{3}{2}(1+\delta)H^{2}\Omega_{m}(t)(1+F)\delta
\end{equation} 
where $F= 0$  reproduces the small-field case and $F =1/3$ corresponds to the large-field limit. It is convenient to write the above equation in terms of $y\equiv \ln a$ (we take the present value of the scale factor $a_0=1$):
\begin{eqnarray}\label{11}
&&H^{2}(y)\frac{d^{2}\delta}{dy^{2}}+ \left[2H^{2}(y)+H(y)\frac{dH(y)}{dy}\right]\frac{d\delta}{dy}-\\
&&-\frac{4H^{2}(y)}{3(1+\delta(y))}\left(\frac{d\delta}{dy}\right)^{2}=\frac{3}{2}(1+\delta)H^{2}(y)\Omega_{m}(y)(1+F)\delta.
\nonumber
\end{eqnarray}

%%%%%%%%%%%%%%%%%%%%%%%%%%%%%%%%%%%%%%%%%%%%%%
\section{Calculating the critical density contrast $\delta_{c}$}
\label{methods}

The critical density can be calculated following two different procedures: one directly from the time evolution of $\delta(y)$ given by Eq.  (\ref{11})  and another from the difference in evolution between the bubble radius and the background scale factor, as we see below. In the former, one has to deal with numerical infinities. We  start with the latter, where one can circumvent this problem. 

%%%%%%%%%%%%%%%%%%%%%%%%%%%%%%%%%%%%%%%%%%%%%%
%%%%%%%%%%%%%%%%%%%%%%%%%%%%%%%%%%%%%%%%%%%%%%
\subsection{The differential-radius method}

We define (following Ref.~\cite{ref36}) the differential radius
\begin{equation}
q (y \equiv \ln a) \equiv \frac{r}{r_{i}}- \frac{a}{a_{i}},
\label{q}
\end{equation}
where $r_{i}$ is the initial bubble radius when its scale factor is $a_{i}$ . The full nonlinear evolution equation for $q(y)$ obtained from mass conservation and Eq.~(\ref{11}) is
\begin{eqnarray}
\label{r4}
q'' + \frac{H'}{H}q' &=& -\frac{1}{2}\frac{\Omega_{m0}a^{-3}- 2\Omega_{\Lambda0}}{\Omega_{m0}a^{-3}+ \Omega_{\Lambda0}}q - \\
&-& \frac{1}{2}\frac{\Omega_{m0}a^{-3}}{\Omega_{m0}a^{-3}+ \Omega_{\Lambda0}}(1+F)\left(\frac{a}{a_{i}}+q\right)\sigma,
\nonumber
\end{eqnarray}
where
\begin{equation}\label{r5}
\sigma \equiv \left(\frac{1}{qa_i/a +1}\right)^3 (1+\delta_i)-1.
\end{equation}
The initial conditions---usually set at a high redshift, when matter dominates ($\Omega_m\sim 1$)---are $ q(y_{i}) \equiv q_{i}=0$ and $q'_{i} = -\delta_{i}(1+p)/(3(1+\delta_i))$, with
\begin{equation}
p \equiv p(F)  \equiv \frac{5}{4}\left(-1 + \sqrt{1 + \frac{24}{25}F}    \right).
\end{equation}
One then solves Eq.~(\ref{r4}) requiring that the collapse, defined by $q(z_c) = -a_c/a_i$, takes place at a given redshift $z_c \equiv 1/a_c -1$. Such requirement constraints $q_i'$ and, consequently, the value of $\delta_i \equiv \delta_m(a_{i})$ which we parametrize as
\begin{equation}
\label{C}
\delta_i = C \, a_i^{1+p},
\end{equation}
inspired by the linear growth of $\delta_c$ at such high redshifts [see Eq.({\ref{linhigh})].

   The critical density $\delta_c$ is the  linear evolution---determined by Eq.~(\ref{b.2})---of such initial density perturbation $\delta_i$, given by Eq.~(\ref{b.8}) at the collapse ($a=a_c$):
$$
\delta_c = C \, a_c^{1+p} \,{_2}F_{1}\left(\epsilon(p),b(p);c(p);-\frac{(1-\Omega_{m0})}{\Omega_{m0}}a_c^{3}\right)
$$
for any $\Omega_{m0}$ and $F$---see Eqs.~(\ref{ctes}) for definitions of $\epsilon(p)$, $b(p)$ and $c(p)$.

The explicit dependence on $\Omega_{m0}$ is usually taken for granted because the full evolution (i.e, until the collapse) is supposed to happen while matter still dominates. In that case there is no dependence of the hypergeometric function on the collapse scale factor $a_c$ for any value of $F$, since ${_2}F_{1}(\epsilon,b;c;0)=1 \; \forall \{\epsilon,b,c\}$ and one recovers the standard power-law dependence
\begin{equation}
\delta_c(a) = C \, a_c^{1+p}.
\label{linhigh}
\end{equation}

Of course, for $p(F=0)=0$ and $\Omega_{m0}=1$, we recover the standard value $\delta_c=1.68647$, as expected. For $F=1/3$ and $\Omega_{m0}=1$, our results yield a constant $\delta_c = 1.70605$ for any $z_c$, which agrees with the result given in Ref.~\cite{ref36} for $z_{c}=0$.

%    We can then determine the value of $C$ defined by Eq.~(\ref{C}) and, accordingly, the value of the critical overdensity as a function of $p(F)$, $z_c$ (or $a_c$) and $\Omega_{m0}$ --- the latter variable being the improvement we present here, for this particular approach:
%
%\begin{equation}
% \delta_c = C \, a_c^{1+p} \,{_2}F_{1}\left(\epsilon(p),b(p);c(p);-\frac{(1-\Omega_{m0})}{\Omega_{m0}}a_c^{3}\right).
%\end{equation}
%
%See Eq.~(\ref{ctes}) for definitions of $\epsilon(p)$, $b(p)$ and $c(p)$.
%Note that the dependence of the Hypergeometric function on the collapse scale factor $a_c$ is discontinued in the EdS case ($\Omega_{m0}=1$) for any value of $F$:
%
%\begin{equation}
%\delta_c(a) = C a_c^{1+p},
%\end{equation}
%
%since ${_2}F_{1}(\epsilon,b;c;0)=1 \, \forall \{\epsilon,b,c\}$. Of course, for $p(F=0)=0$ and $\Omega_{m0}=1$ we recover the standard value $\delta_c=1.68647$, as expected. For $F=1/3$ and $\Omega_{m0}=1$, our results yield a constant $\delta_c = 1.70605$ for any $z_c$, which agrees with the result given in Ref.~\cite{ref36} for $z_{c}=0$.

%%%%%%%%%%%%%%%%%%%%%%%%%%%%%%%%%%%%%%%%%%%%%%%%%%%
%%%%%%%%%%%%%%%%%%%%%%%%%%%%%%%%%%%%%%%%%%%%%%%%%%%
\subsection{The constant-Infinity method and its mending}

As mentioned before, the full solution of Eq.~(\ref{11}) should be infinite at the collapse. However, since this equation can be solved only numerically, it is necessary to establish the infinite as a very large number for a given collapse redshift $z_ c$. As we see below, this numerical infinity depends both on the collapse redshift $z_c$ and on $\Omega_{m0}$.

%\subsubsection{Initial conditions} 
Here, we consider an initial value for the redshift $z_{i}=1000$, where the Universe is completely dominated by nonrelativistic matter ($\Omega_{m}=1 $). We then have to deal with the linear version of Eq. (\ref{11}) when the background behaves like an EdS universe:
\begin{equation}\label{12}
\frac{d^{2}\delta}{dy^{2}}+ \frac{1}{2}\frac{d\delta}{dy}=\frac{3}{2}(1+F)\delta_{m},
\end{equation}
whose linear growing solution is given by
\begin{equation}\label{13}
\delta_l(y)=C e^{y(1+p)},
\end{equation}
where $C$ is again a constant which clearly depends on the collapse redshift.
The initial conditions are then given by
\begin{equation}
\label{15}
\delta_{i}= C e^{y_{i}(1+p)}\quad ,\quad
\delta'_{i}= (1+p)\delta_{i}.
\end{equation}
The constant $C$ is obtained requiring the collapse occurs at $y=y_c$, i.e, $\delta(y_c)\to \infty$. To fix the numerical value of infinity, we use the known values of $\delta_{c}$  for an EdS universe: 
\begin{equation}
\label{dtheo}
\delta_c=
\left\{	
	\begin{array}{lcl}
	\frac{3}{5}\left(\frac{3\pi}{2}\right)^{2/3} &,& {\rm for \quad}  F=0	\\
	~ & ~ \\
	1.70605 &,& {\rm for \quad}  F=1/3,	
	\end{array}
\right.
\end{equation}
where the latter value was obtained in the previous section. The constant $C$ is then given by
\begin{equation}
\label{16}
C(y_c)=
\left\{	
	\begin{array}{ccl}
	\frac{3}{5}\left(\frac{3\pi}{2}\right)^{2/3} e^{-y_c}&,& {\rm for \quad}  F=0	\\
	~ & ~ \\
	1.70605 \,  e^{-y_c(1+p)}&,& {\rm for \quad}  F=1/3.	
	\end{array}
\right.
\end{equation}

Evolving  Eq. (\ref{11}) with the initial conditions (\ref{15}) and (\ref{16}), the numerical infinity is defined by
\begin{equation}
\label{17}
{\rm Inf}(y_c)\equiv \delta_{m}(\Omega_{m0}=1, y_c),
\end{equation}
which clearly depends on $z_c$. Some papers \cite{Ronaldo1,infinito} do not take this dependence into account and assume a constant value $\widetilde{\rm Inf} \equiv {\rm Inf}(y_c=0) = 10^5$ or $10^8$.
%%%V2
While both $\widetilde{\rm Inf}$ and ${\rm Inf}$ have roughly the same order of magnitude ($\sim 10^5$) at $z_c = 0$, ${\rm Inf}(z_c)$ decreases monotonically with $z_c$ and can be as low as $10^4$ at $z \sim 3$ (for both $F=0$ and $F=1/3$).
%%%
Later on (see Figs. \ref{error}, \ref{error11}, and \ref{errorNbin_13-14}), we point out the numerical differences in the final outcome from this approximation. 

Once the so-called infinity ${\rm Inf}(y_c)$ is established, we then use Eq. (\ref{11})  for different values of collapse redshifts $z_c$ and for different values of $\Omega_{m0}$ in order to calculate the constants $C_{j}\equiv C(y_c,\Omega_{m0})$  that satisfy the condition for the collapse  $\delta_{m}(\Omega_{m0},y_c)= {\rm Inf}(y_c)$. 
The corresponding value of $\delta_{c}$ is given by Eq. (\ref{b.8}). If $F=0$,  we get
\begin{eqnarray}
\label{19}
\delta_{c}(\Omega_{m0},y_c) &=& C_{j} \, e^{y_c} \times \\
&\times& {_2}F_{1}\left[\frac{1}{3},1;\frac{11}{6};-\frac{(1-\Omega_{m0})}{\Omega_{m0}}e^{3y_c}\right].\nonumber
\end{eqnarray}
If $F=1/3$,  we get
\begin{eqnarray}
\label{5.16}
\delta_{c}(\Omega_{m0},y_c)&=& C_{j} \, e^{\frac{1}{4}(-1+\sqrt{33})y_c} \times \\
&&\hspace{-0.14 \textwidth} \times  {_2}F_{1}\left[\frac{7+\sqrt{33}}{12}, \frac{-1+\sqrt{33}}{12};\frac{6+\sqrt{33}}{6};-\frac{(1-\Omega_{m0})}{\Omega_{m0}}e^{3y_c}\right] . \nonumber
\end{eqnarray}

%%%%%%%%%%%%%%%%%%%%%%%%%%%%%%%%%%%%%%%%%%%%%%%
\section{Comparing the results from different approaches}
\label{conclusion}

In this section we plot the relative errors between our method and the previously mentioned ones, for both values $F=0$ and $F=1/3$, when one calculates the critical density $\delta_c$ (as a function of the collapse redshift), the comoving number density of halos per logarithmic mass interval $n_{\ln M}$ [see Eq.~(\ref{10.6} for the definition], and the number of clusters at a given redshift in  mass bins $N_{\rm bin}$, as a function of either the redshift or the mass interval.

If the value of $z_{c}$ is fixed, our method describes the evolution of $\delta_{c}$  in terms of $\Omega_{m0}$ (Fig. \ref{dOm}, left panel). On the other hand, when we fix the value of $\Omega_ {m0}$, the method describes the evolution of $\delta_{c}$ as function of redshift $z_ {c}$ (Fig. \ref{dOm}, right panel). 
\begin{figure}
\includegraphics[width=0.47\textwidth]{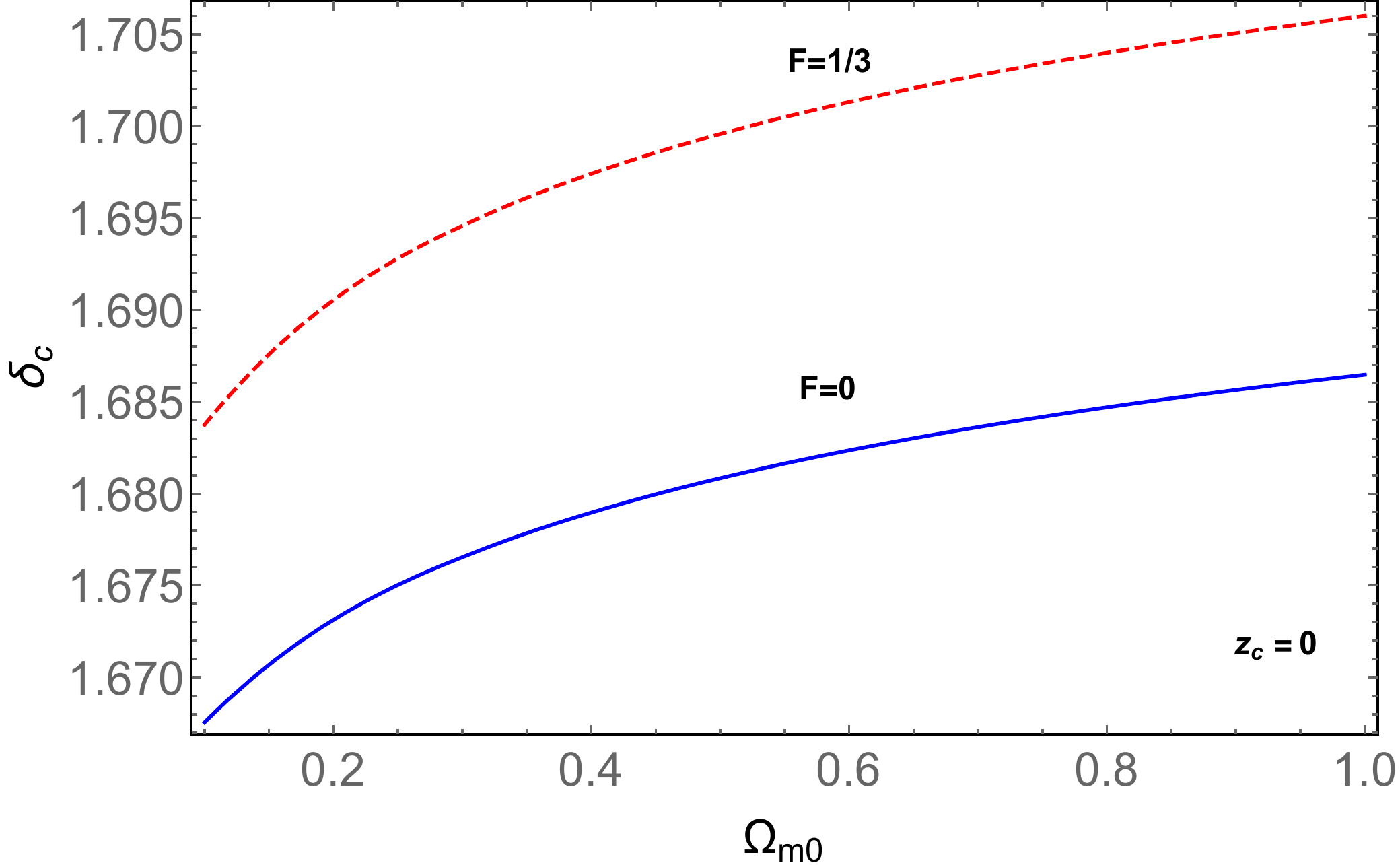}
\includegraphics[width=0.47\textwidth]{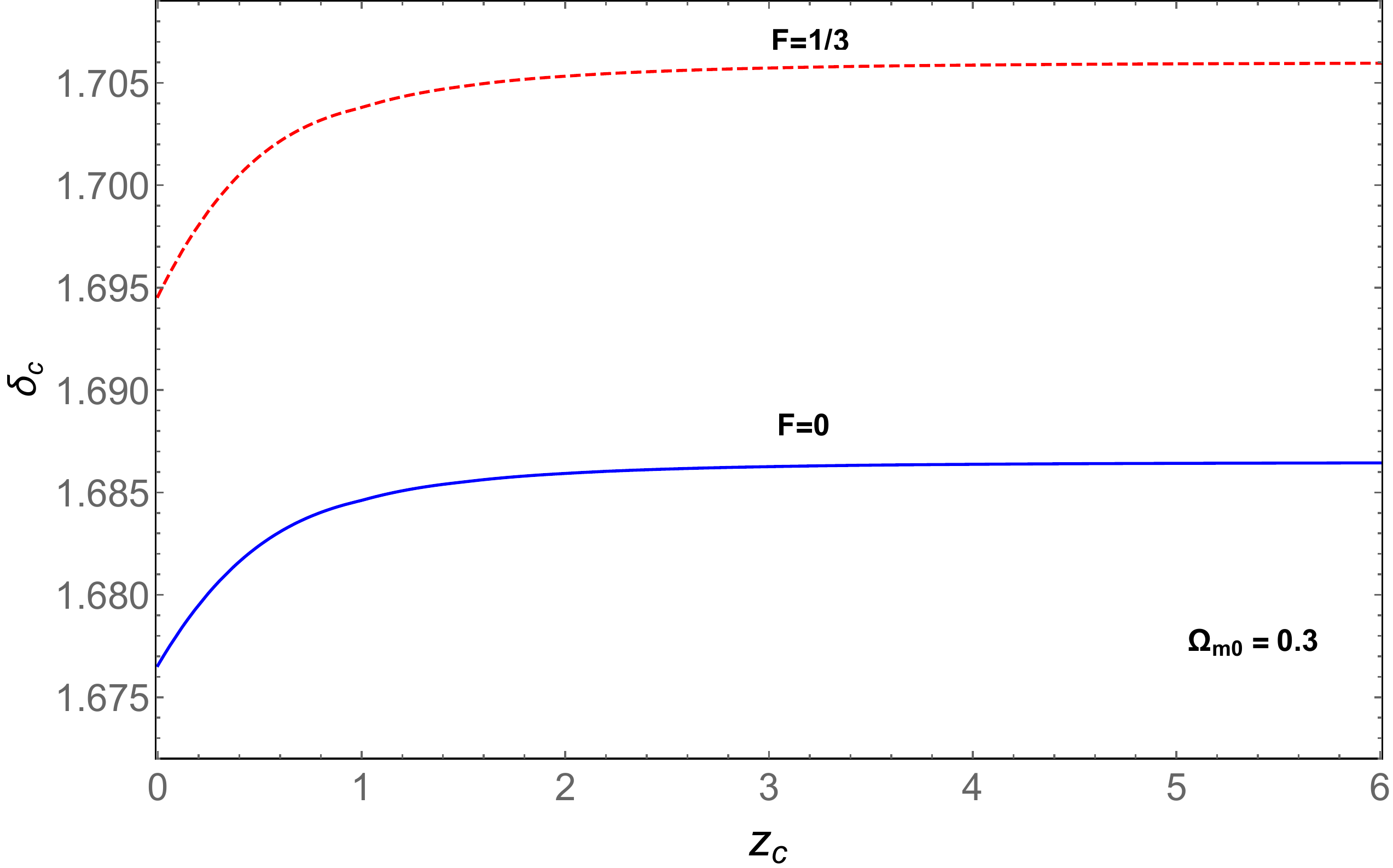}
\centering
\caption{(a,left panel) Critical density contrast $\delta_{c}$ as a function of $\Omega_{m0}$ for $F=0$  (solid blue line) and $F=1/3$ (dotted red line) when the collapse occurs at redshift $z_c=0$, following our approach. (b,right panel) Critical density contrast $\delta_{c}$ as function of redshift collapse $z_{c}$ for $F=0$  (solid blue line) and $F=1/3$ (dotted red line) with $\Omega_{m0}=0.3$, following our approach.} 
\label{dOm}
\end{figure}
%
%\begin{figure}
%\includegraphics[width=0.6\textwidth]{Figura2new.pdf}
%\centering
%\caption{Critical density contrast $\delta_{c}$ as function of redshift collapse $z_{c}$ for $F=0$  (solid blue line) and $F=1/3$ (dotted red line) with $\Omega_{m0}=0.3$, following our approach.} 
%\label{dz}
%\end{figure} 
%

In the following figures we compare the results of our method with the differential-radius equation and with the method where the numerical infinity assumes large constant values (we pick $\widetilde{\rm Inf}=10^{5}$  and $10^{8}$ for the sake of comparison with previous results in the literature \cite{Ronaldo1,infinito}) for both values $F=0$ and $F=1/3$. We define the ratio between other method and our as

\begin{equation}
\Delta^{i} \equiv \frac{ \delta_c^{i} } { \delta_c^{\rm pw} }
\end{equation}
where the superscript $i$ stands for the different methods in the literature, specified in the figures, and  ${\rm pw}$ stands for the approach introduced in the present work.
For the figures we always assume $\Omega_{m0}=0.3$.

In Fig \ref{error} we show $1-\Delta^i$ as a function of the collapse redshift $z_c$. 
Note that the differential-radius method and ours are equivalent. On the other hand, the relative differences between our method (or the differential-radius method) and the constant-infinity one increase (in magnitude) with the collapse redshift because the critical density contrast calculated with our method stabilizes at the EdS value for large $z_c$ (as it is supposed to), while it grows unbounded in the latter approaches. 
\begin{figure}
\includegraphics[width=0.47\textwidth]{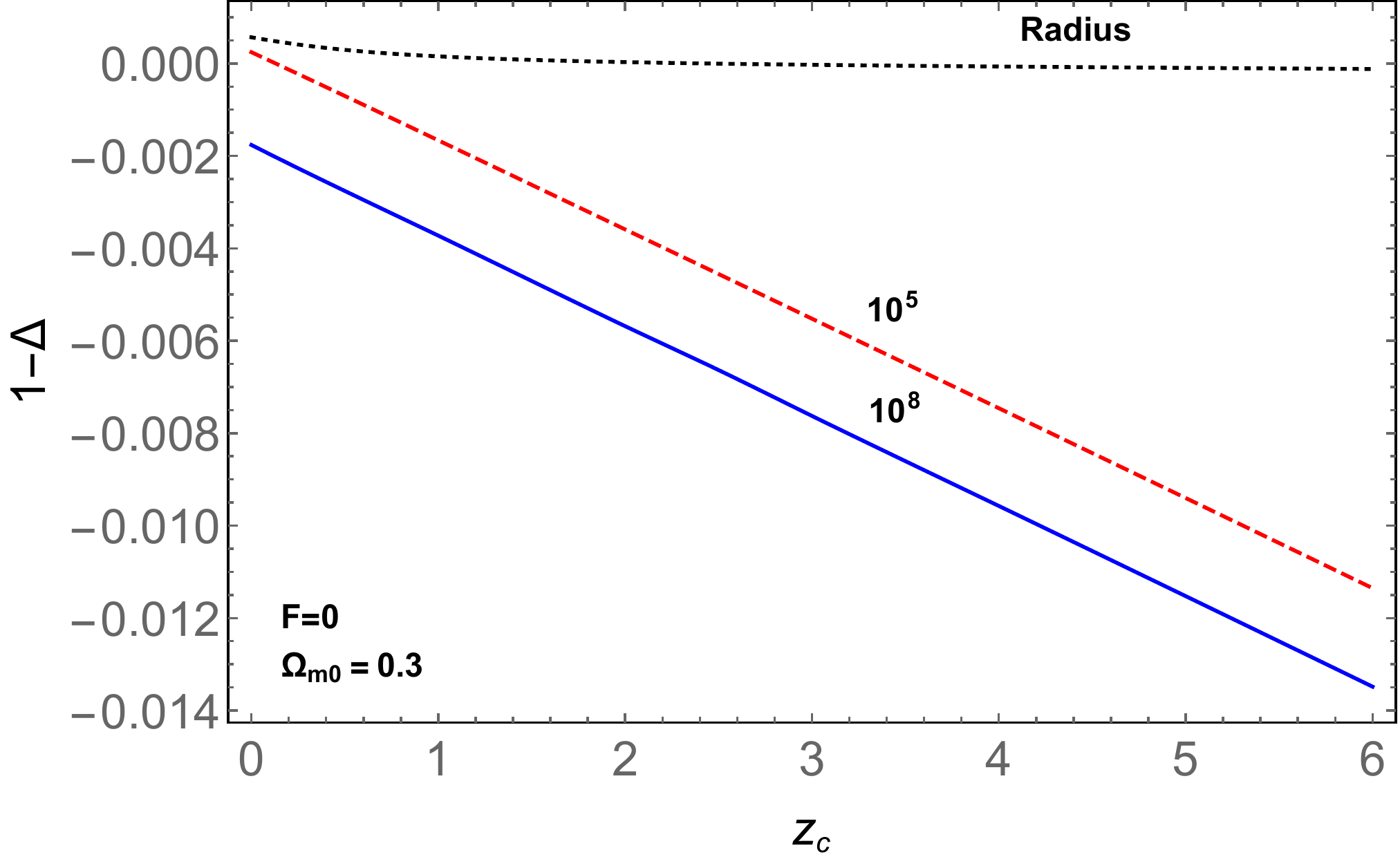}
\includegraphics[width=0.47\textwidth]{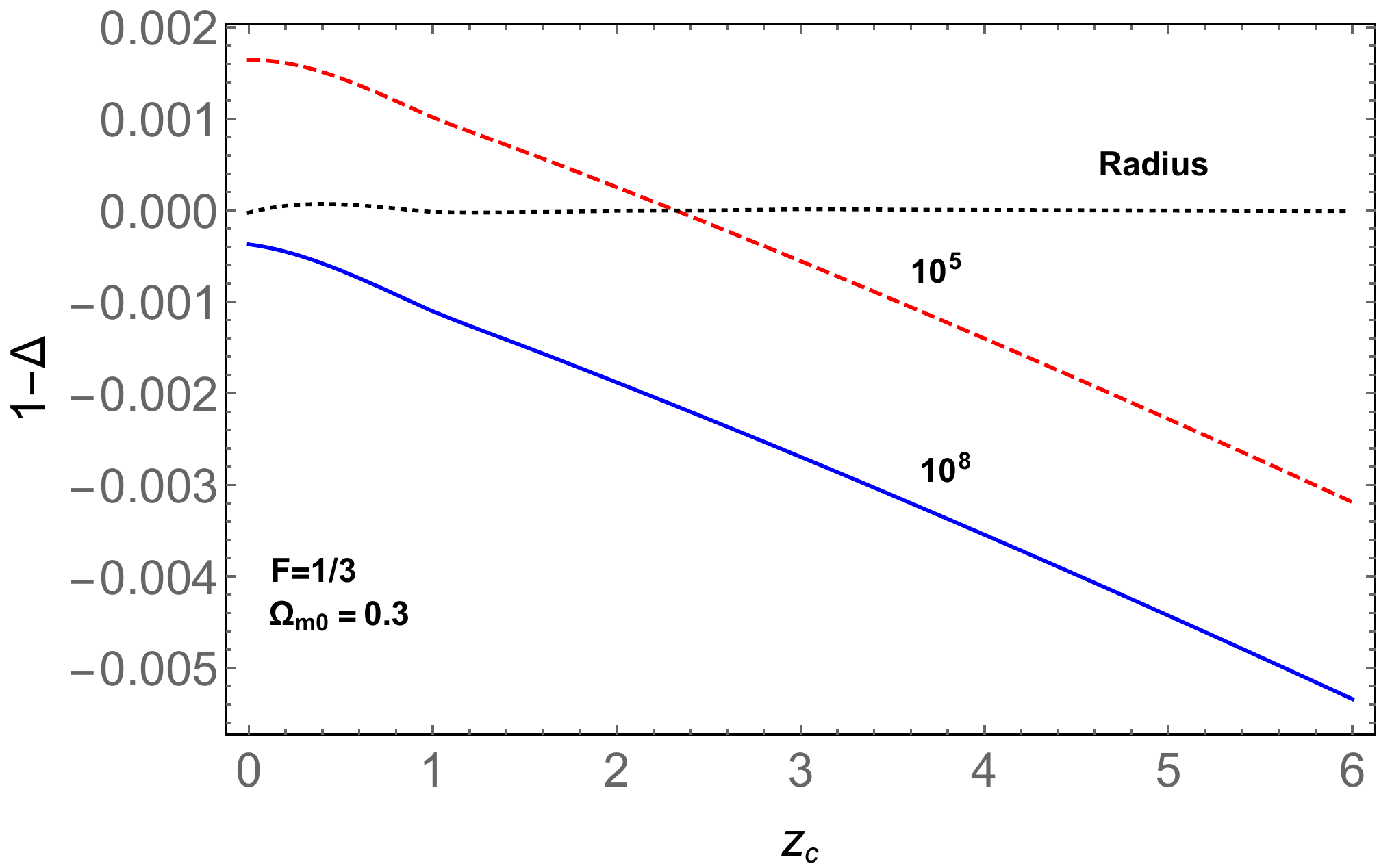}
\centering
\caption{Relative errors for $\delta_c$ (see text for definition) between our method and {\bf (a)} the differential-radius one (dotted black line), the constant-infinity one, with  {\bf (b)} $\widetilde{\rm Inf} =10^{5}$ (dashed red line) and {\bf (c)} $\widetilde{\rm Inf} =10^{8}$ (solid blue line) for $F=0$ (left panel) and $F=1/3$ (right panel). For all models, we assumed $\Omega_{m0}=0.3$.
}\label{error}
\end{figure}

Next, we calculate and compare the mass function using the results  for $\delta_{c}$ obtained by our method and by the method where the infinite is fixed  (we pick $10^{5}$ and $10^{8}$ again). We use the Sheth-Tormen mass function  \cite{Sheth}, which analytically determines the distribution of these objects as a function of their virial masses and redshifts. 
%%%V2
The virial mass $M$ is defined such that the average density inside the virial radius $r_v$ is $\Delta_v$ times the critical density. We should mention that in order to relate SC to virialized halos, the virial theorem has to be modified in $f(R)$ theories. See, for instance, Ref.~\cite{Schmidt} for more on this topic.”   
%
%By assuming a NFW form \cite{navarro}, one could rescale the the virial mass $M$ to $M_{300}$ --- more often used in simulations (see, for instance, Refs.~\cite{ref36,Hu_Kravtsov}).
%%%
In this formalism, the comoving number density of halos per logarithmic interval (in the virial mass $M$) is
\begin{equation}\label{10.6}
n_{\ln M}=\frac{dn}{d \ln M}(M,z)=\frac{\rho_{m0}}{M}f(\nu)\frac{d\nu}{d \ln M}
\end{equation}
where $\rho_{m0}$ is the present matter density of the Universe, the peak threshold $\nu= \delta_{c}(z)/\sigma(M,z)$,
\begin{equation}
f(\nu)= A \sqrt{\frac{2}{\pi}a\nu}[1+ (a\nu^{2})^{-q}]\exp [-a\nu^{2}/2)]
\end{equation}
and $\sigma$  is the variance of the linear contrast density field in spheres of a comoving radius $r$ containing the mass $M$.Also, $A$ is a normalization constant such that $\int f(\nu)d\nu=1$. Here, we use the following approximation \cite{Viana} for $\sigma$:
\begin{equation}\label{13.6}
\sigma(M,z)= \sigma_{80} \, D(z)\left(\frac{M}{M_{8}}\right)^{-\gamma(M)/3}, 
\end{equation}
where
\begin{equation}\label{14.6}
\gamma(M)=(0.3 \Gamma + 0.2)\left[2.92 + 
\frac{1}{3}\log\left(\frac{M}{M_{8}}\right)\right],
\end{equation}
and shape parameter $\Gamma$ is given by \cite{Viana}
\begin{equation}\label{15.6}
\Gamma= \Omega_{m0}h \,\mbox{exp} \left(-\Omega_{b}-\frac{\Omega_{b}}{\Omega_{m0}}\right).
\end{equation}
Above, $M_{8}=5.95$ $\mbox{x}$ $ 10^{14}\Omega_{m0}h^{-1} M_{\odot}$ is  the mass inside a sphere of radius $8h^{-1}\mbox{Mpc}$, $h$ is  the current Hubble constant in units of $100 {\rm ~km~ s}^{-1} {\rm Mpc}^{-1}$, $\Omega_{b}=0.02230/h^{2}$  is the baryonic  density parameter, and
$D(z)$  is the growing solution of Eq.~(\ref{b.0}) normalized in $z=0$. We assume that $\sigma_{80}=0.8159$ \cite{ref19}, $q=0.3$, and $a=0.707$ \cite{Sheth}. We have checked that if, instead of the fit given by Eq. (\ref{13.6})---that is numerically simpler to deal with---we had used a more accurate approach, our results would not change significantly.

In Fig.  \ref{error11}, we show  the relative error
\begin{equation}
1-\Delta_{n}^i \equiv \left[n_{\ln M}^{\rm pw}- n_{\ln M}^i\right]/n_{\ln M}^{\rm pw}
\end{equation}
for $z=0$  between our method and the others when $\Omega_{m0}=0.3$ for both values $F=0$ and $F=1/3$.
\begin{figure} 
\includegraphics[width=0.47\textwidth]{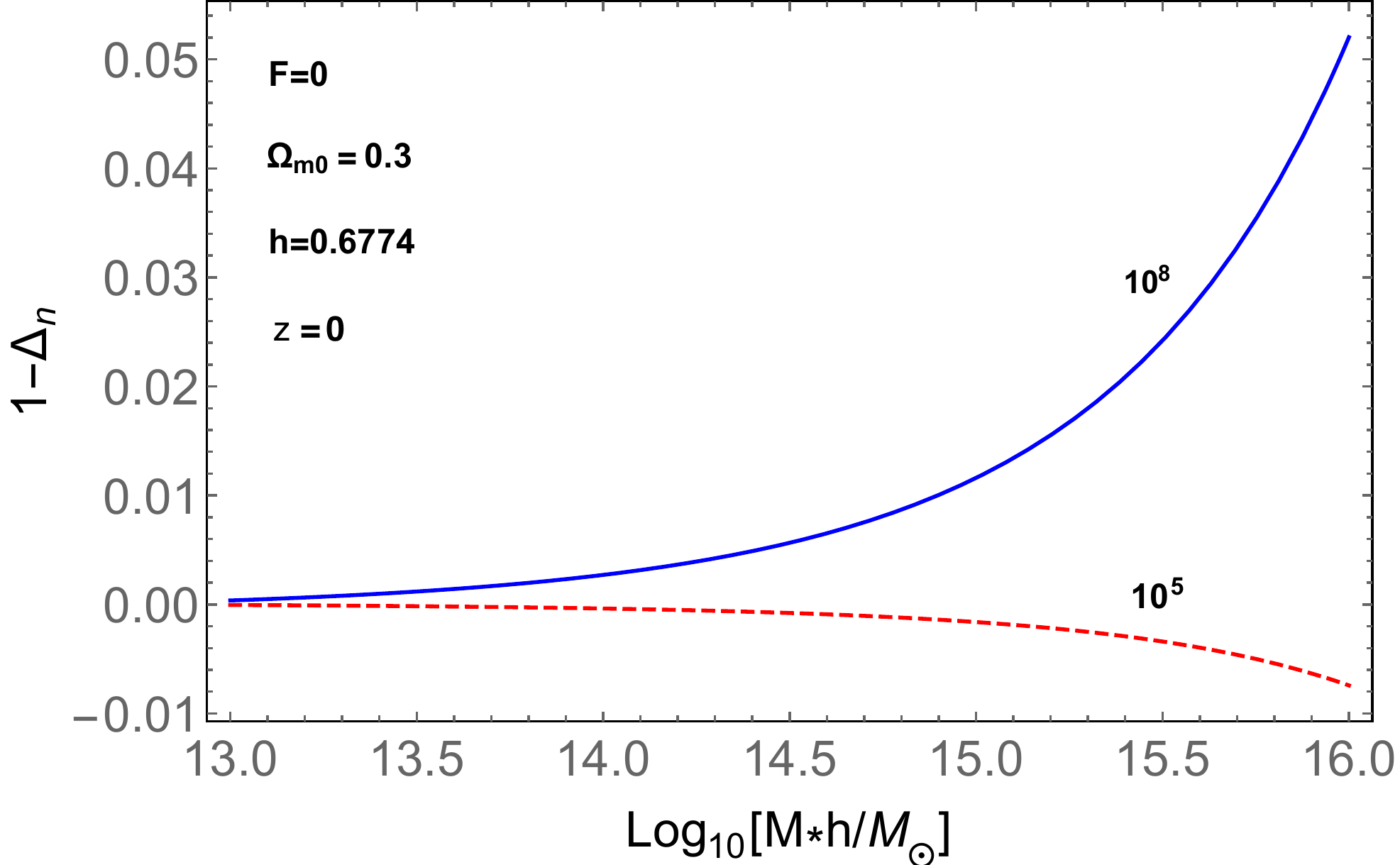}
\includegraphics[width=0.47\textwidth]{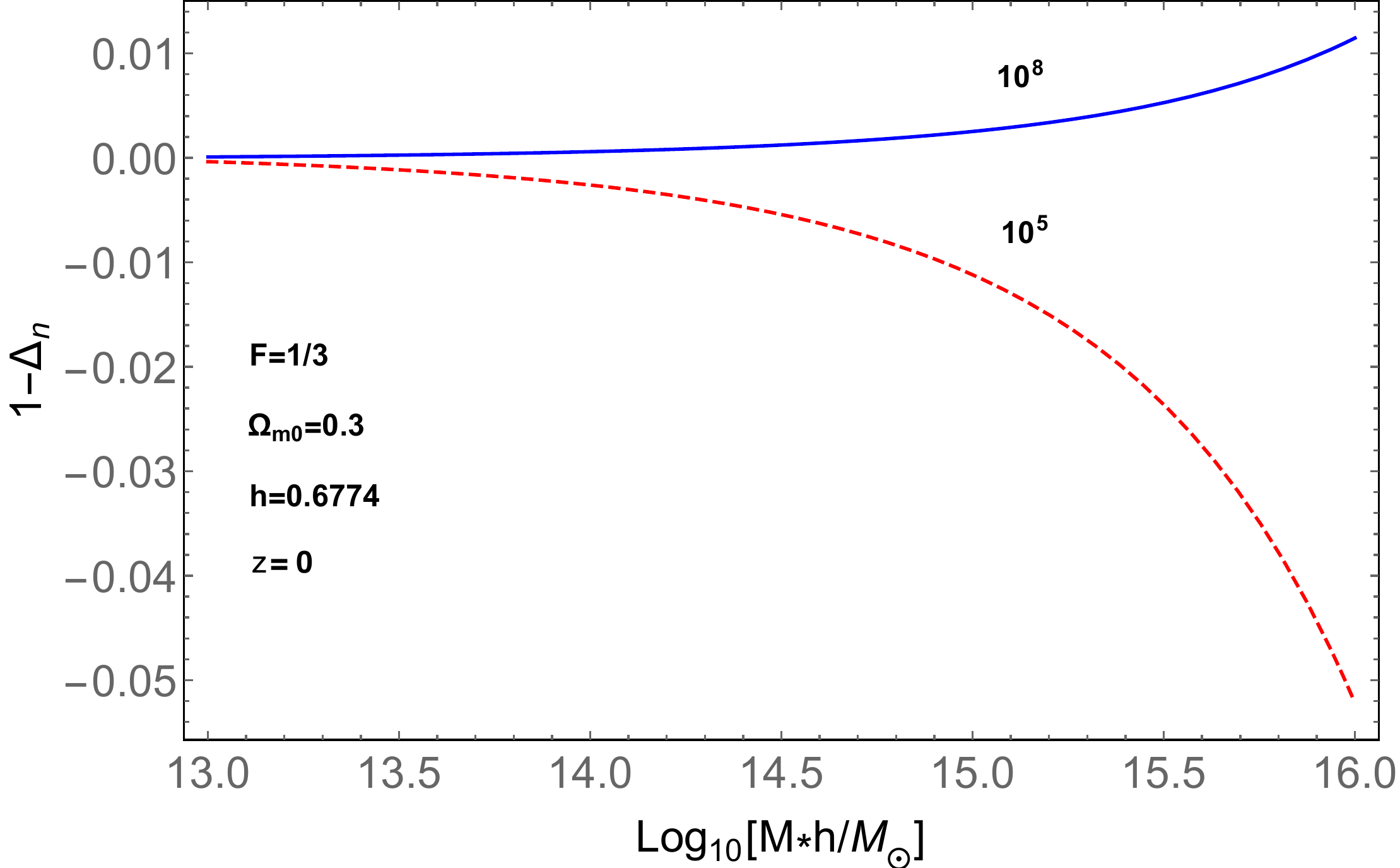}
\centering
\caption{Relative errors for $n_{\ln M}$ (see text for definition) between our method and the constant-infinity one, with  {\bf (a)} $\widetilde{\rm Inf} =10^{8}$ (solid blue line) and {\bf (b)} $\widetilde{\rm Inf} =10^{5}$ (dashed red line) for redshift $z=0$ in the cases $F=0$ (left panel) and $F=1/3$ (right panel). For all models, we assumed $\Omega_{m0}=0.3$ and $h=0.6774$.}
\label{error11}
\end{figure}

In Fig.  \ref{error111} we show relative errors as a function of the redshift when the virial mass takes  the value  $10^{13}h^{-1}M_{\odot}$ (upper panels) and $10^{15}h^{-1}M_{\odot}$ (lower panels). From these figures, it is clear that the relative errors are more significant for larger virial masses and higher redshift.
\begin{figure} 
\includegraphics[width=0.47\textwidth]{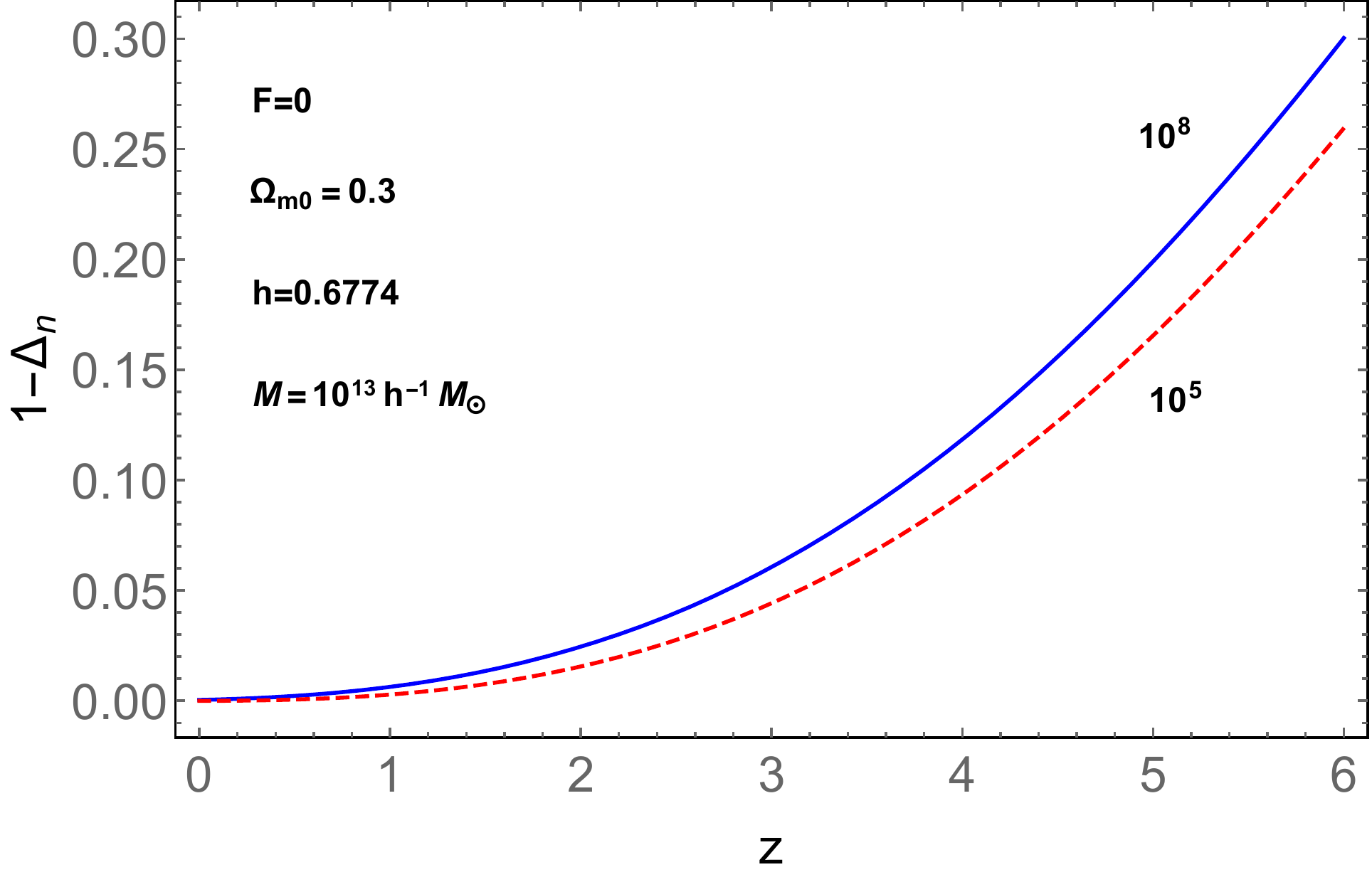}
\includegraphics[width=0.47\textwidth]{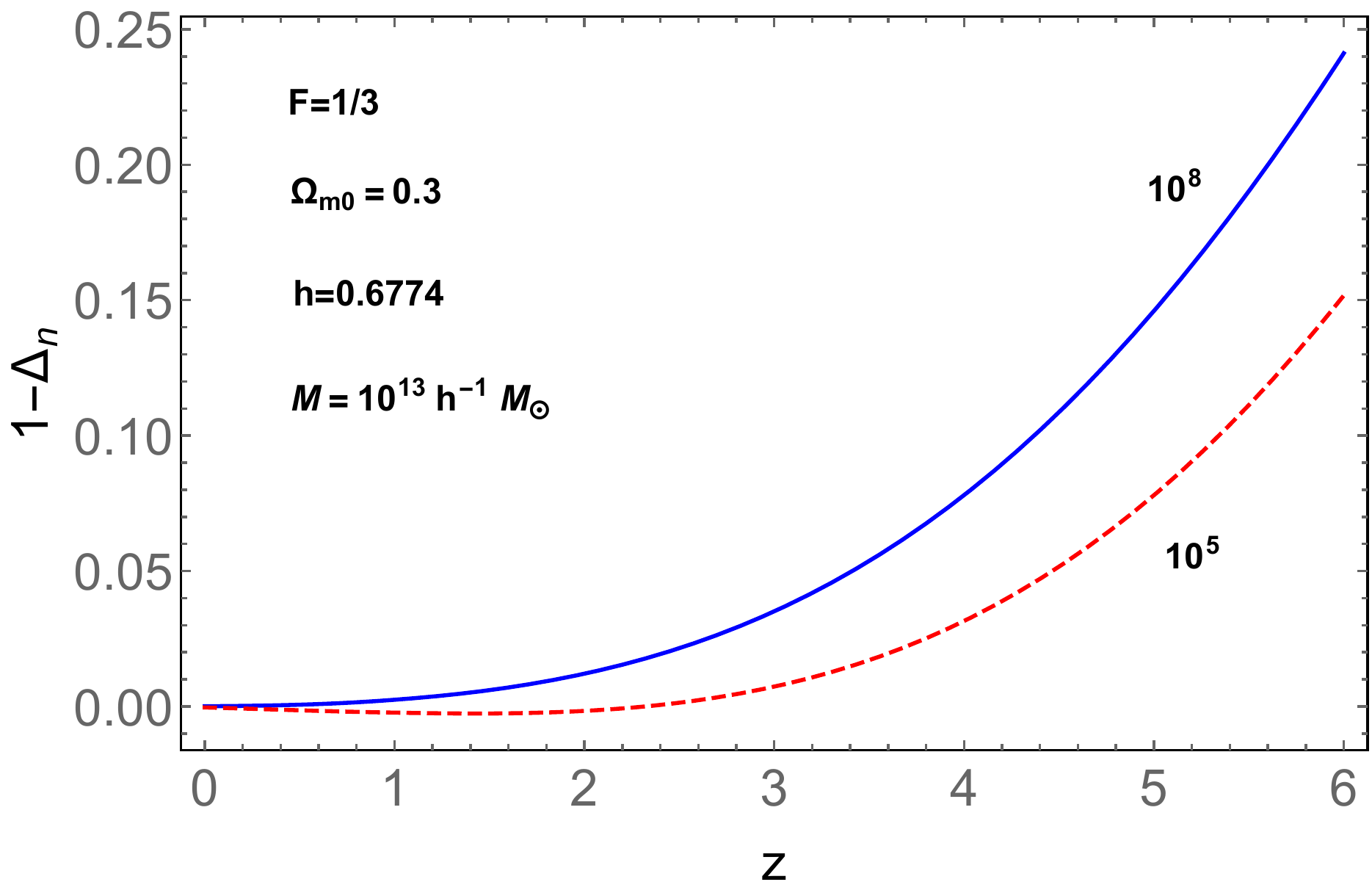}
\includegraphics[width=0.47\textwidth]{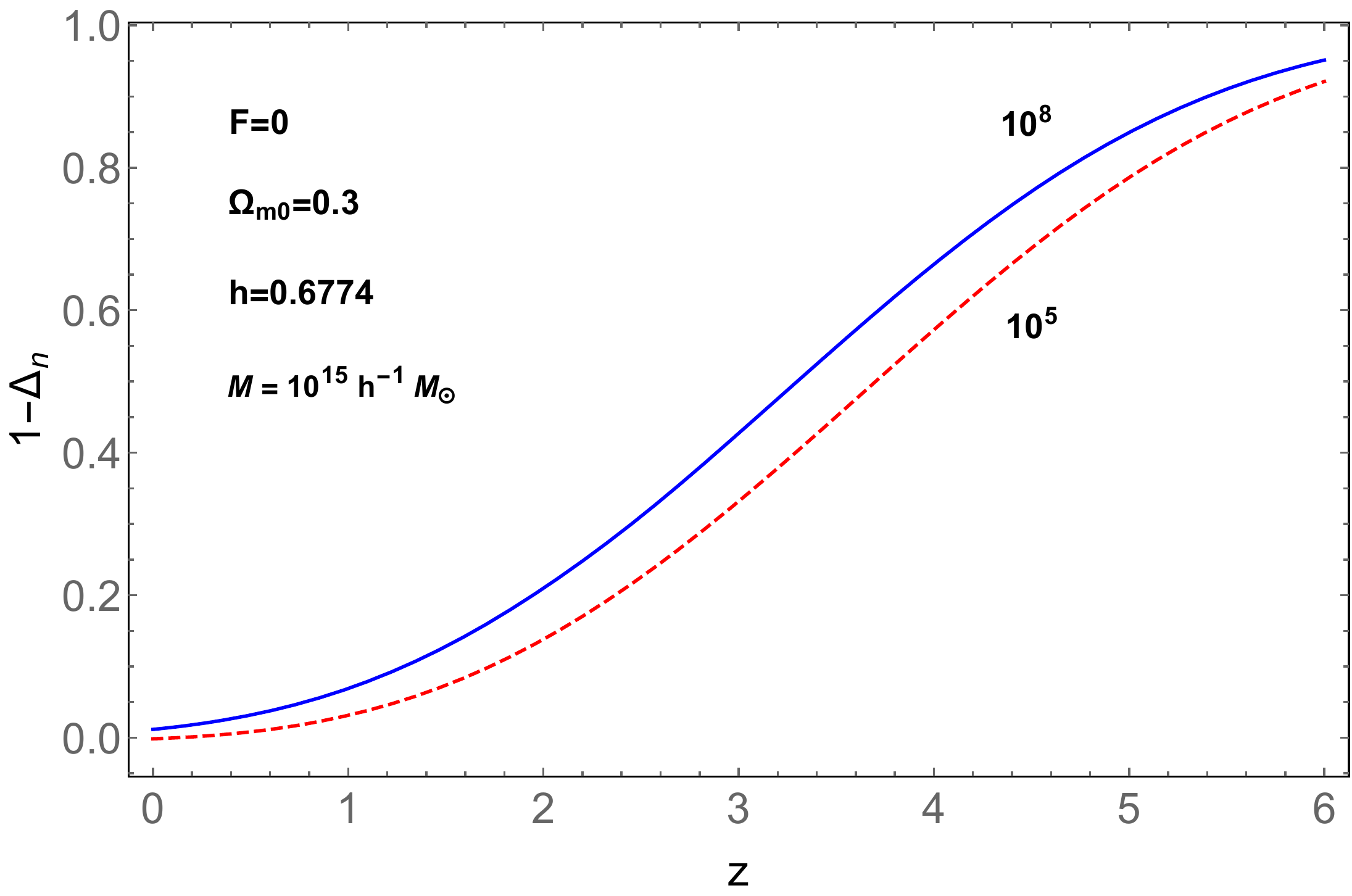}
\includegraphics[width=0.47\textwidth]{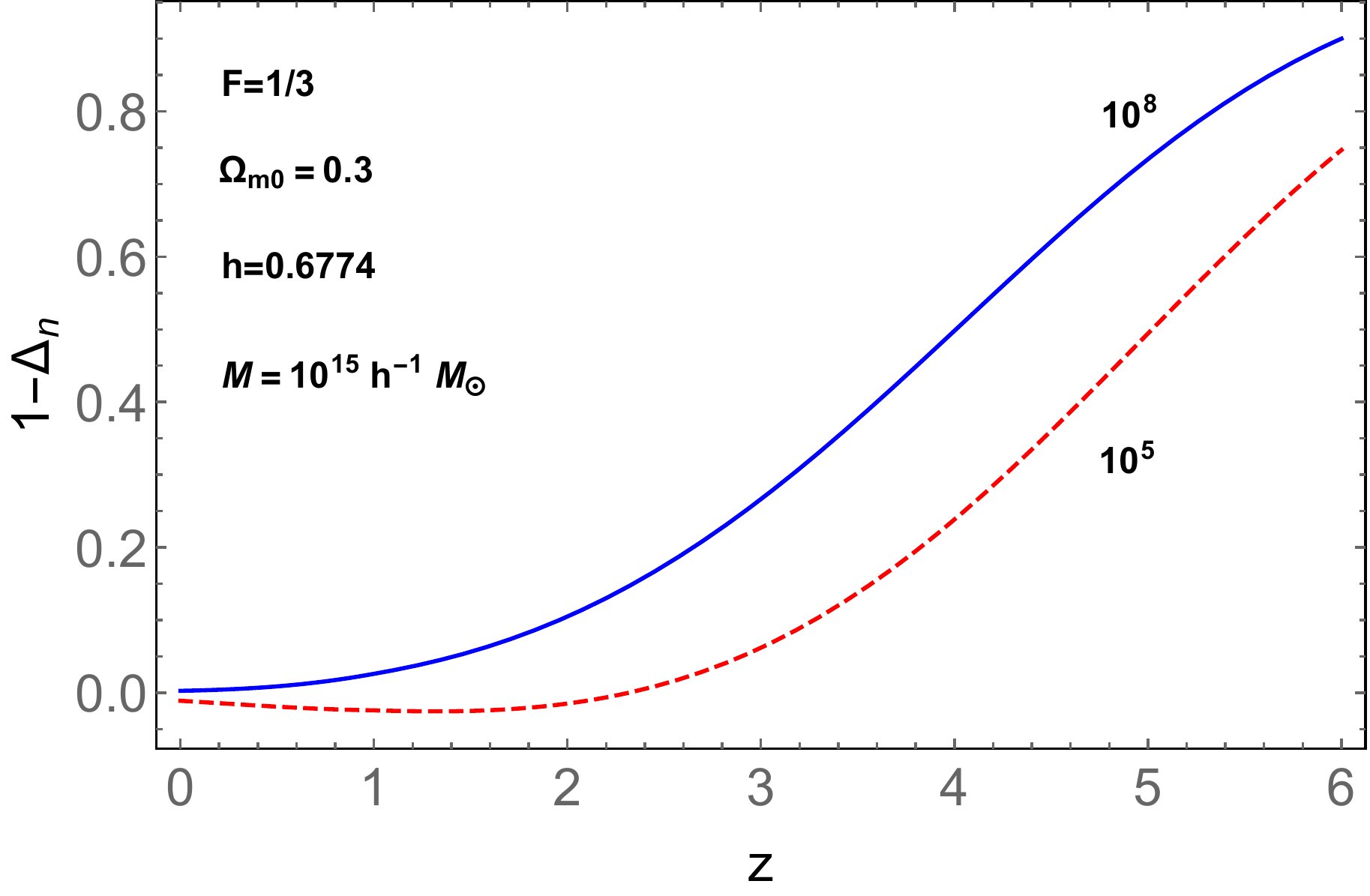}
\centering
\caption{Relative errors  for $n_{\ln M}$ (see text for definition) between our method and the constant-infinity one, with  {\bf (a)} $\widetilde{\rm Inf} =10^{8}$ (solid blue line) and {\bf (b)} $\widetilde{\rm Inf} =10^{5}$ (dashed red line) as a function of redshift for $F=0$, $F=1/3$ and  virial mass of  $10^{13}h^{-1}M_{\odot}$ (upper panels). The same plots, are shown for $10^{15}h^{-1}M_{\odot}$ (lower panels). For all models, we assumed $\Omega_{m0}=0.3$ and $h=0.6774$.}
\label{error111}
\end{figure}

The observed quantity, however, is the number of clusters at a given redshift and in a given mass bin. It is defined as
\begin{equation}
N_{\rm bin} \equiv \int_{4\pi} d\Omega \int_{M_{\rm inf}}^{M_{\rm sup}} \frac{dn_{\ln M}}{dV}\frac{dV}{dz \, d\Omega} \, dM,
\end{equation}
where $dV$ is the comoving volume at redshift $z$, $\frac{dV}{dz \, d\Omega} = \frac{r^2}{H(z)}$, and the comoving distance $r(z)$ is given by
\begin{equation}
r(z) = \int_0^z H^{-1}(z') \, dz'.
\label{comoving}
\end{equation}
Figure~\ref{Nbin_13-14}, obtained with our method, shows $N_{\rm bin}$ for $F=0$ and virial masses in the ranges $10^{13}-10^{14}$ $ h^{-1}M_{\odot}$ (left panel) and $10^{14}-10^{15}$ $ h^{-1}M_{\odot}$ (right panel). Similar results are obtained by using $F=1/3$. Note that for higher-mass bins, the peak in the $N_{\rm bin}$ is lower and located at a smaller $z$. From this piece of information and from our results above---both $\Delta^i$ and $\Delta_n^i$ increase with $z$---we expect the bin $10^{13}-10^{14}$ $ h^{-1}M_{\odot}$ to be the most sensitive one to the small differences in $\delta_c$ and $n_{\ln M}$ we have been pointing out.
\begin{figure} 
\includegraphics[width=0.47\textwidth]{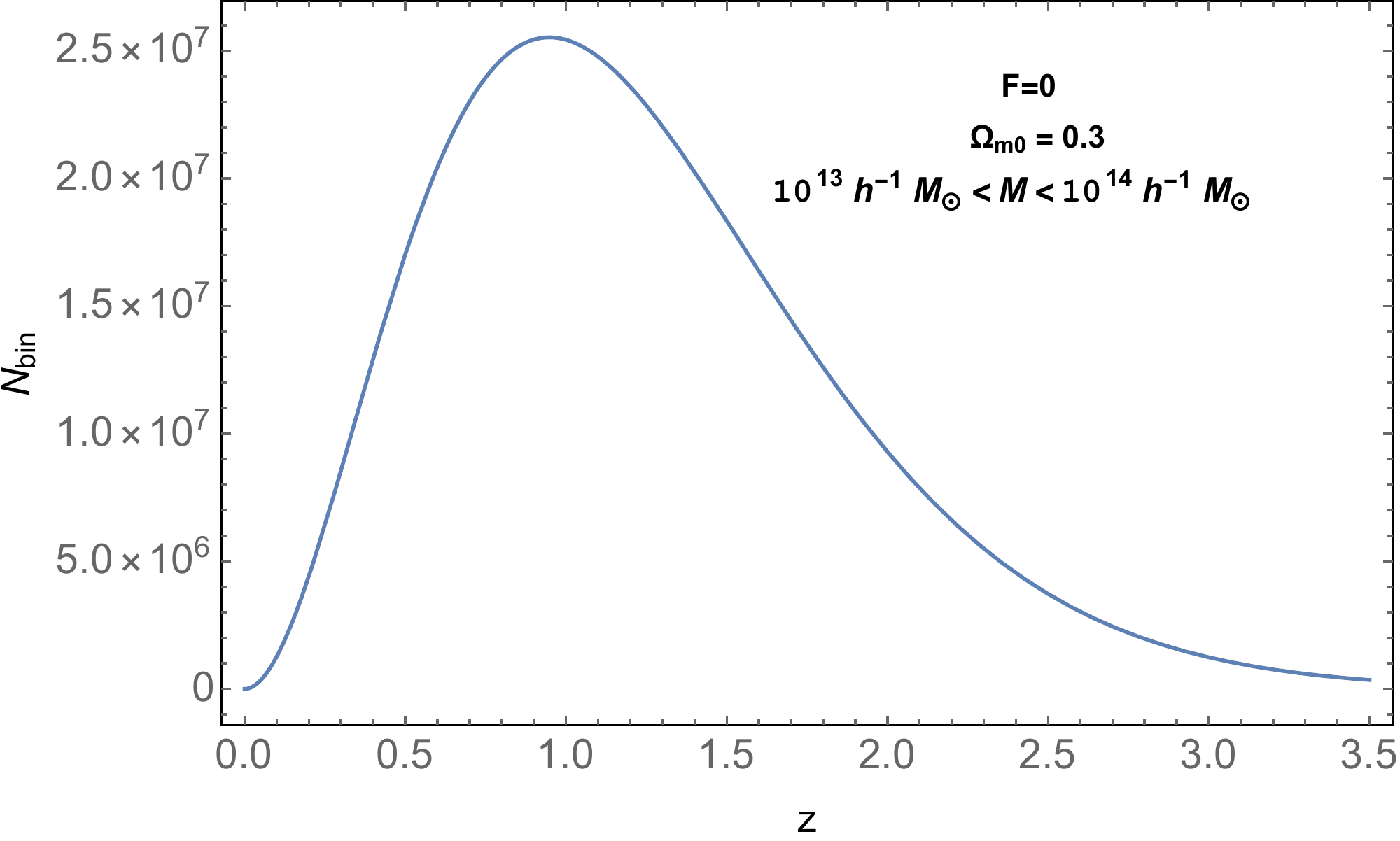}
\includegraphics[width=0.47\textwidth]{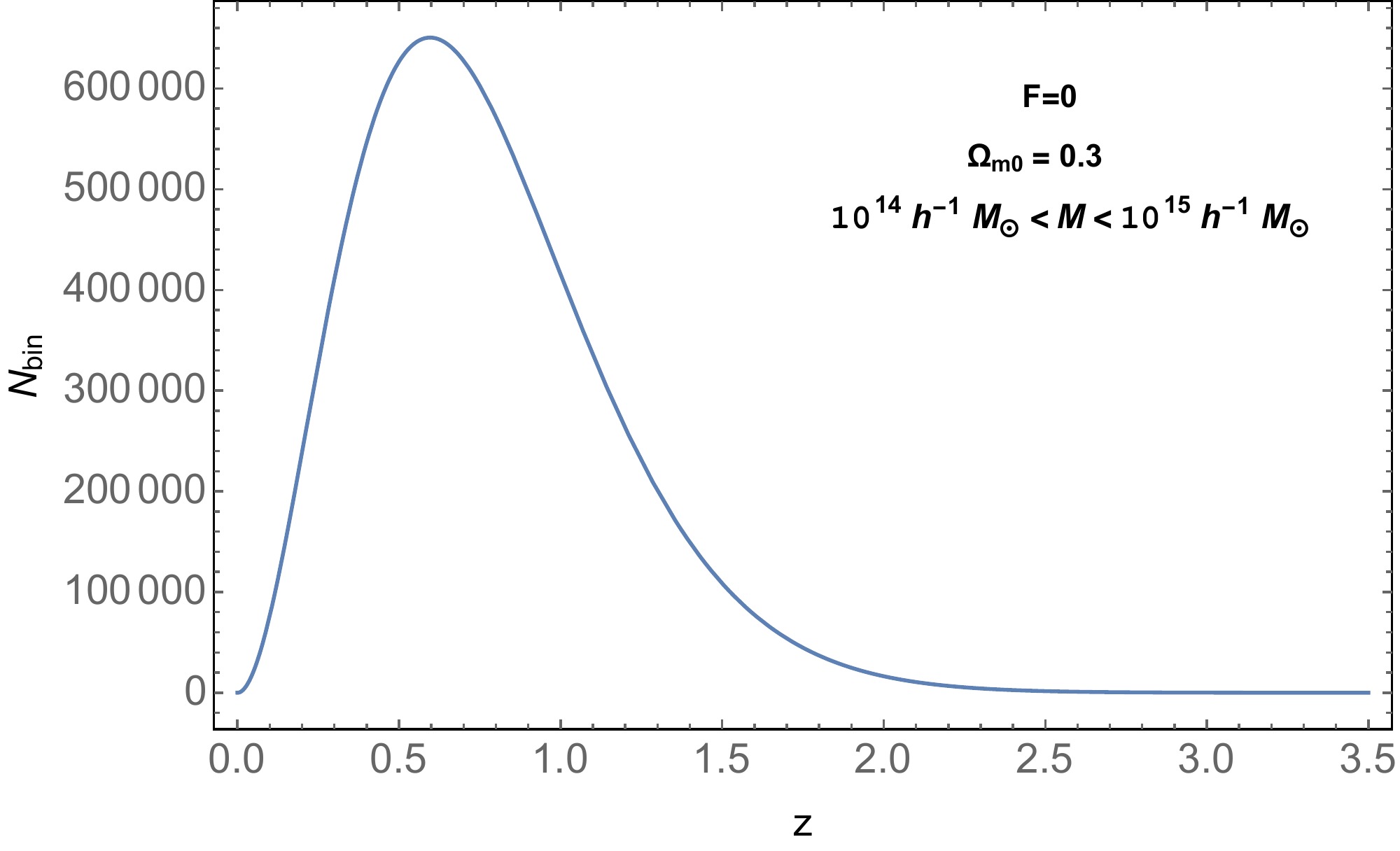}
\centering
\caption{$N_{\rm bin}$ (see text for definition) as a function of redshift for $F=0$ and virial masses between $10^{13}$ and $10^{14}$ $ h^{-1}M_{\odot}$ (left panel) and $10^{14}$ and $10^{15}$ $ h^{-1}M_{\odot}$ (right panel). As always, we assumed $\Omega_{m0}=0.3$ and $h=0.6774$.}
\label{Nbin_13-14}
\end{figure}
We then define one last relative error, namely,
\begin{equation}
1-\Delta_{N_{\rm bin}}^i \equiv \left[N_{\rm bin}^{\rm pw} - N_{\rm bin}^i\right]/N_{\rm bin}^{\rm pw}
\end{equation}
and plot it, for virial masses between $10^{13}$ and $10^{14}$ $ h^{-1}M_{\odot}$, in Fig.~\ref{errorNbin_13-14} for $F=0$ (left panel) and $F=1/3$ (right panel). Notice that, in the redshift range of interest, it is at most of the order of $0.01$. 
%Therefore, we conclude that, although the approach presented here is more rigorous than the ``constant-infinity'' method used in the literature, at least at the current stage of the observations, it does not yield observable differences in the cluster number count.
%
\begin{figure} 
\includegraphics[width=0.47\textwidth]{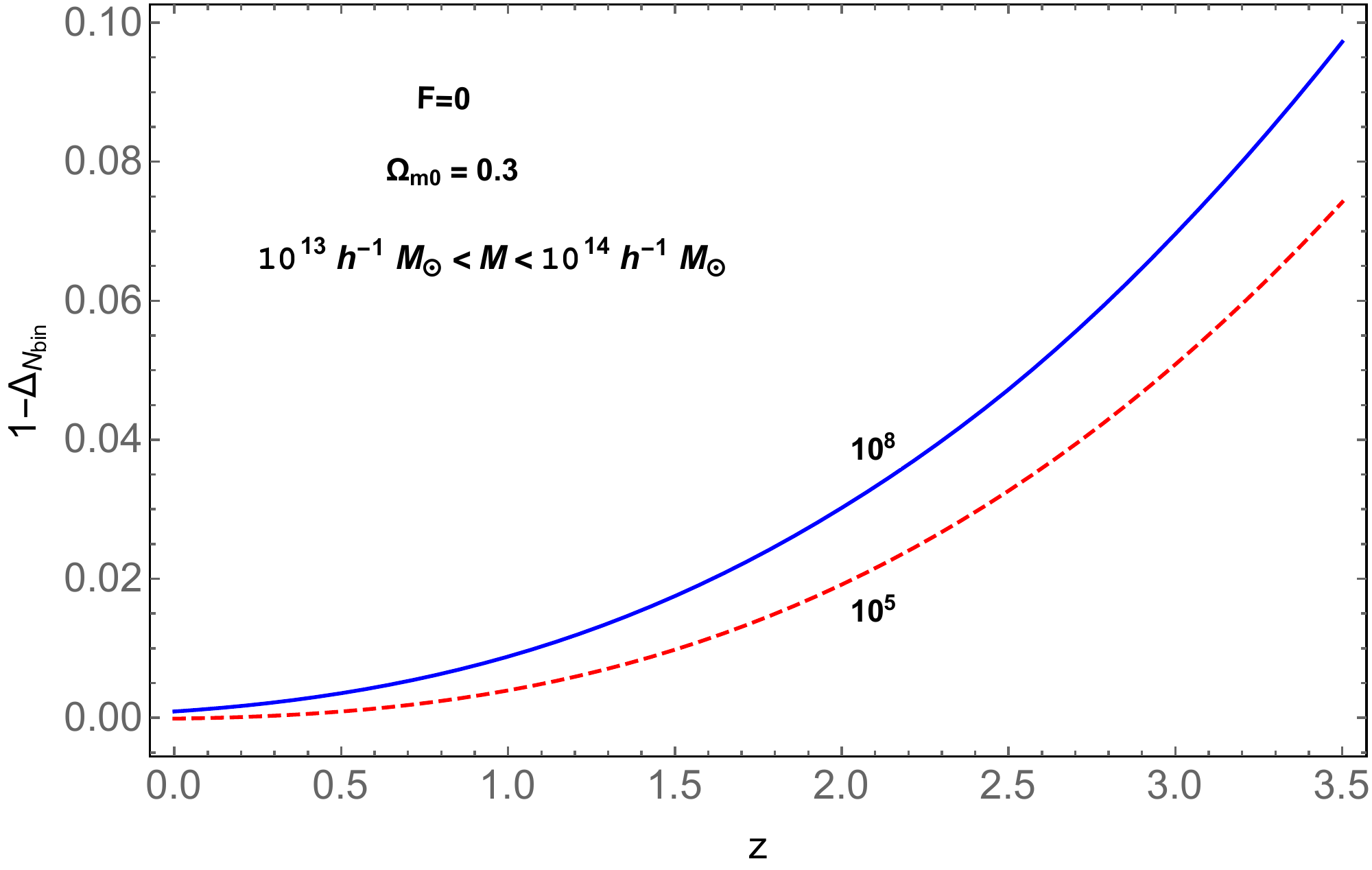}
\includegraphics[width=0.47\textwidth]{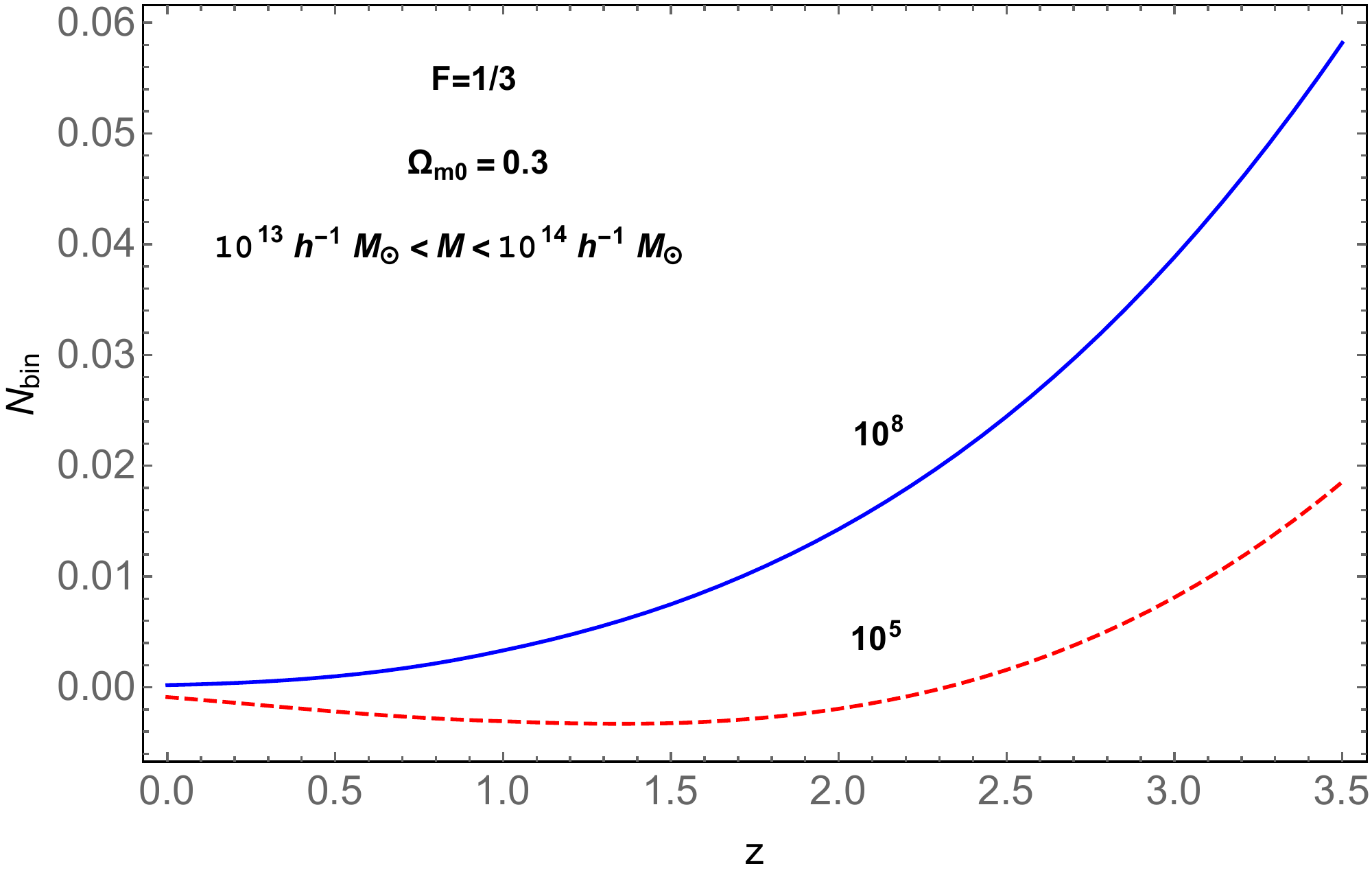}
\centering
\caption{Relative errors for $N_{\rm bin}$ (see text for definition) between our method and the constant-infinity one, with  {\bf (a)} $\widetilde{\rm Inf} =10^{5}$ (dashed red line) and {\bf (b)} $\widetilde{\rm Inf} =10^{8}$ (solid blue line) as a function of redshift for virial masses between $10^{13}$ and $10^{14}$ $ h^{-1}M_{\odot}$ and $F=0$ (left panel) and $F=1/3$ (right panel). For all models, we assumed $\Omega_{m0}=0.3$ and $h=0.6774$.}
\label{errorNbin_13-14}
\end{figure}
%%%%%%%%%%%%%%%%%%%%%%%%%

\section{Conclusions}
\label{conclusions}

In summary, we have shown that our approach matches the results from the differential-radius method and pointed out that the so-called constant-infinity method does need corrections in the calculation of a key quantity, namely, the critical density $\delta_c$. 
%In particular, in $F(R)$ gravity theories, we were able to recover the standard analytical limit of EdS. 
We have also derived an analytical expression for the critical density $\delta_c$ as a function of $\Omega_{m0}$, $z_c$, and $F$.

%To the best of our knowledge, this is the first calculation of $N_{\rm bin}$ with such robust method. 

In spite of being more rigorous and more accurate than the constant-infinity method used in the literature, we should mention that, for the the current stage of the observations, the procedure presented here does not yield observable differences in the cluster number count $N_{\rm bin}$.  The small discrepancies pointed out (less than $1\%$) may be of use in the future, when more precise data become available. 

%%%V2
We remark that our results are based on the validity of the SC approximation and of the Sheth and Tormen prescription. It would be interesting to compare the differences we found, due to distinct ways of calculating $\delta_c$, using a semianalytic approach, with those from N-body simulations. Besides, as shown in Ref.~\cite{Burrage}, departures from spherical symmetry affect chameleon screening and a detailed comparison of semianalytical methods and simulations are required to determine the correct functional form of the mass function. This is an important task to have in mind in the upcoming large-scale surveys.

\section*{\textbf{ACKNOWLEDGEMENTS}}
D.H. acknowledges financial support from CAPES.

%%%%%%%%%%%%%%%%%%%%%%%
\appendix
\section{SOLUTION OF THE DIFFERENTIAL
 EQUATION FOR THE DENSITY CONTRAST IN
 THE LINEAR REGIME}
\label{App}

Consider the linear equation for the perturbations
\begin{equation}\label{b.0}
 \ddot{\delta} + 2H\dot{\delta}- \frac{3}{2}(1+F)\Omega_{m}(t)H^{2}(t)\delta=0,
\end{equation}
where $\dot ~\equiv d/dt$. Considering the change of variable  $t\rightarrow a(t)$ and using the equation for the background in the standard $\Lambda$CDM scenarium (neglecting radiation), namely,
\begin{equation}\label{b.1}
 H^{2}=H_{0}^{2}(\Omega_{m0}a^{-3}+(1-\Omega_{m0})),
\end{equation}
then Eq. (\ref{b.0}) is written as
\begin{eqnarray}
\label{b.2} 
 a^{2}\delta''&+& \left(3-\frac{3/2\Omega_{m0}}{\Omega_{m0}+(1-\Omega_{m0}a^{3})}\right)a\delta' - \\
 &-& \frac{3}{2}(1+F)\Omega_{m}(a)\delta=0, \nonumber
\end{eqnarray}
where $' \equiv d/da$. In an EdS universe, the above expression becomes
\begin{equation}\label{b.3}
 a^{2}\delta''+ \left(\frac{3}{2}\right)a\delta'-\frac{3}{2}(1+F)\delta=0.
\end{equation}
Suppose a solution to (\ref {b.3}) of the form
\begin{equation}
\label{b.4}
 \delta= C a^{1+p},
\end{equation}
where $ C $ is a constant. Replacing this solution in Eq. (\ref{b.3}) yields
\begin{equation}
\label{b.5}
 a^{p+1}\left[ p(1+p)+\frac{3}{2}(1+p)-\frac{3}{2}(1+F)\right]=0,
\end{equation}
whose nontrivial solution is given by
\begin{equation}
 p=-\frac{5}{4} \pm \frac{5}{4}\sqrt{1+\frac{24}{25}F}.
\end{equation}

Now consider a solution to Eq. (\ref{b.2}) of the form
\begin{equation}
 \delta_{m} \propto a^{1+p}G(a).
\end{equation}
Substituting this solution in Eq. (\ref{b.2}), we get
\begin{eqnarray}
\label{b.7}
a^{2}G'' &+& \left(5+2p -\frac{3}{2}\Omega_{m}(a)\right)aG'+ \\
 &&\hspace{-0.1\textwidth}+ (1+p)\left[(p+3)-\frac{3}{2}\Omega_{m}(a)\left(1+\frac{1+F}{1+p}\right)\right]G=0.\nonumber
\end{eqnarray}
Making the change of variable
\begin{equation}
 u(a)\equiv -\frac{(1-\Omega_{m0})}{\Omega_{m0}}a^{3},
\end{equation}
one can recognize Eq. (\ref{b.7}) as an hypergeometric differential equation. 
Thus, the growing solution of Eq. (\ref{b.2}) is
\begin{equation}
\label{b.8}
 \delta_{m} \propto a^{1+p}\,{_2}F_{1}(\epsilon,b;c;u),
\end{equation}
where ${_2}F_{1}(\epsilon,b;c;u)$ is the hypergeometric function,  and
\begin{eqnarray}
 \label{ctes}
 \epsilon \equiv \epsilon(p) &\equiv&\frac{1}{3}\left[(2+p) + \sqrt{(2+p)^{2}-(p+3)(p+1)}\right], \nonumber \\
 b\equiv b(p) &\equiv&\frac{(p+3)(1+p)}{3\left[(2+p) + \sqrt{(2+p)^{2}-(p+3)(p+1)}\right]},  \\
 c\equiv c(p) &\equiv&\frac{1}{3}\left(7+2p-\frac{3}{2}\right).\nonumber
 \end{eqnarray}
For $F=0$,  one obtains the solution \cite{ioav}
\begin{equation}
\delta_{m} \propto a \, {_2}F_{1}\left[1 ,\frac{1}{3};\frac{11}{6};-\frac{(1-\Omega_{m0})}{\Omega_{m0}}a^{3}\right],
\end{equation}
and for $F=1/3$, we get
\begin{eqnarray}
\delta_{m}&\propto& a^{(-1+\sqrt{33})/4} \times \\
&& \hspace{-0.1\textwidth} \times
{_2}F_{1}\left[\frac{7+\sqrt{33}}{12} ,\frac{\sqrt{33}-1}{12};\frac{6+\sqrt{33}}{6};-\frac{(1-\Omega_{m0})}{\Omega_{m0}}a^{3}\right].
\nonumber
\end{eqnarray}
%%%%%%%%%%%%%%%%%%%

\bibliographystyle{unsrt}
\bibliography{biblio}

\end{document}